\documentclass[sigchi]{acmart}

\usepackage{booktabs} 
\usepackage[normalem]{ulem} 
\usepackage{xcolor} 

\copyrightyear{2019} 
\acmYear{2019} 
\setcopyright{none} 
\acmPrice{15.00}
\acmISBN{978-1-4503-5970-2/19/05}
\acmDOI{10.1145/3290605.3300432}

\definecolor{copper}{rgb}{0.72, 0.45, 0.2}

\begin{document}
\title{Decision-Making Under Uncertainty in Research Synthesis: Designing for the Garden of Forking Paths}

\author{Alex Kale}
\affiliation{%
  \institution{University of Washington}
  \city{Seattle}
  \state{Washington}
}
\email{kalea@uw.edu}

\author{Matthew Kay}
\affiliation{%
  \institution{University of Michigan}
  \city{Ann Arbor}
  \state{Michigan}
}
\email{mjskay@umich.edu}

\author{Jessica Hullman}
\affiliation{%
  \institution{Northwestern University}
  \city{Evanston}
  \state{Illinois}
}
\email{jhullman@northwestern.edu}

\renewcommand{\shortauthors}{Kale, Kay, \& Hullman}
\renewcommand{\shorttitle}{Decision-Making Under Uncertainty in Research Synthesis}

\begin{abstract}
To make evidence-based recommendations to decision-makers, researchers conducting systematic reviews and meta-analyses must navigate a \textit{garden of forking paths}: a series of analytical decision-points, each of which has the potential to influence findings. To identify challenges and opportunities related to designing systems to help researchers manage uncertainty around which of multiple analyses is best, we interviewed 11 professional researchers who conduct research synthesis to inform decision-making within three organizations. We conducted a qualitative analysis identifying 480 analytical decisions made by researchers throughout the scientific process. We present descriptions of current practices in applied research synthesis and corresponding design challenges: making it more feasible for researchers to try and compare analyses, shifting researchers' attention from rationales for decisions to impacts on results, and supporting communication techniques that acknowledge decision-makers' aversions to uncertainty. We identify opportunities to design systems which help researchers explore, reason about, and communicate uncertainty in decision-making about possible analyses in research synthesis.
\end{abstract}

%
%

\ccsdesc[500]{Information systems~Decision support systems}
\ccsdesc[500]{Human-centered computing~Visualization theory, concepts and paradigms}

\keywords{Representing Uncertainty, Research Synthesis}


\maketitle

\section{Introduction}
Organizations routinely rely on prepared summaries of empirical evidence to support decision-making. For example, the Navy employs scientists who review scientific literature and collect data internally in order to recommend improvements in training practices.
Similarly, Veterans' Affairs employs researchers who meta-analyze the scientific literature on treatments for post-traumatic stress disorder (PTSD) and other conditions in order to recommend the best possible treatment options for veterans struggling with trauma. 

When compiling and communicating a summary of scientific evidence, researchers make a series of analytical decisions such as how to combine information from studies conducted with different measures or in different settings.
Recent work in reproducible statistics~\cite{Silberzahn2018,Simmons2011,Simonsohn2015,Steegen2016,Wicherts2016}, driven by concerns about a ``replication crisis'', demonstrates how flexibility in decision-making produces multiple possible sequences of analytical decisions, which an analyst chooses between at their discretion.
In the context of research synthesis, alternative possible analyses may lead to alternative understandings of empirical evidence and consequently, opposite or inconclusive recommendations. 
Faced with a ``garden of forking paths''~\cite{gelman2014}, researchers cannot eliminate subjectivity and uncertainty from the scientific process.
Instead, scholars suggest that researchers should attempt to understand which analytical decisions impact results~\cite{Simonsohn2015,Steegen2016}.

By identifying researcher degrees of freedom~\cite{Wicherts2016} as a cause of the ``replication crisis'', prior work seems to focus blame on the indiscretions of individual researchers. 
However, with existing software to tools, it is difficult for researchers to deliberate about and explore the consequences of alternative analyses (e.g.,~\cite{Simonsohn2015,Steegen2016}),
such that even researchers with honest intentions may struggle to perceive the implications of different choices.
In this study, we seek an in-depth understanding of how researchers make analytical decisions in research synthesis, and where they struggle with uncertainty in the process, in order to identify opportunities to design 
for decision-making in the garden of forking paths.
We contribute the results of 
a qualitative analysis of open-ended conversational interviews we conducted with 11 researchers who work to support evidence-based decision-making at three institutions: the Navy, the Medical Center at a large public university, and the Veterans' Affairs Medical Center in a major US city. 
In our interviews we elicited detailed descriptions of experiences conducting scientific review and analysis, emphasizing the reasoning behind analytical decisions and the strategies that researchers use to manage uncertainty.
Based on these interviews, we identify a set of challenges around managing uncertainty in research synthesis: the tension between surveying analysis paths and implementing a specific path, the disconnect between researchers' rationales for analytical decisions and the actual impact of those decisions on findings, and the balance between researchers' skepticism and the need for compelling recommendations.
We draw on utility theory~\cite{vonNeumann1944} and prior work on reliable statistics and uncertainty visualization, in combination with our interviews, to identify opportunities to address these challenges by designing systems which encourage exploration of alternative analyses, elicit and represent researchers' reasoning about analytical decisions, and provide researchers with techniques to communicate uncertainty in the research process.
\section{Background}

\subsection{Managing Uncertainty in Decision-Making}

A large body of research on judgment and decision-making (JDM) has examined how people reason with and make decisions under uncertainty.
Canonical work by Tversky and Kahneman~\cite{Kahneman2011,tversky1975} established that people often seek to reduce uncertainty, sometimes by substituting heuristic judgments for more complex reasoning.
A drive to reduce uncertainty can lead to unwarranted expressions of certainty~\cite{Manski2018}, which has consequences for decision-making individually and at an organizational level (e.g., in public policy).


Decision-making under uncertainty is characterized by feelings of conflict and doubt which block or delay a choice between alternative courses of action~\cite{March1958}.
As such uncertainty is broadly defined and associated with a variety of terms~\cite{Boukhelifa2017,Lipshitz1997} such as ambiguity~\cite{Hogarth1987,March1976}, risk~\cite{Anderson1981,Arrow1965,MacCrimmon1986}, unreliability, imprecision, incompleteness, and contradiction~\cite{Klir1994}, as well as error and subjectivity~\cite{MacEachren2005}.
Based on a literature review of work describing real world decision-making under varying forms of uncertainty, Lipshitz and Strauss~\cite{Lipshitz1997} developed a framework for understanding how decision-makers in organizations like the military cope with uncertainty. In their framework uncertainty is either \textit{acknowledged} through preemptive action and planning, \textit{reduced} through rule- or assumption-based reasoning, or \textit{suppressed} through ignoring information or guesswork. 
We borrow these strategies from Lipshitz and Strauss to characterize analytical decisions described by our participants during interviews.


\subsection{Systematic Review and Meta-Analysis}
Systematic review and meta-analysis are methodologies used to produce a rigorous summary of existing evidence on a topic. 
Using systematic review, researchers account for scientific literature within a consistent framework and characterize the research to date on a particular topic~\cite{Nelson2014}. 
As an extension of systematic review, researchers sometimes choose to aggregate quantitative results from studies through a meta-analysis~\cite{Cooper2009,Lipsey2001,Nelson2014}. Meta-analysis produces an estimate of the effect size of an intervention by pooling statistical outcomes from studies conducted under similar conditions. 

Ideally, when prescribed procedures are followed, systematic review and meta-analysis offer more robust findings~\footnote{See Manski's critique~\cite{Manski2018-lure} about difficulty interpreting meta-analyses.} than regular literature review and quantitative analysis, respectively~\cite{Nelson2014}.
A typical systematic review starts with a question about the effect of some intervention and database queries to find all the relevant literature.
Researchers independently
judge which articles to include and exclude from the review and then resolve disagreements. 
Since working in pairs is standard practice, this is sometimes called dual-review.
For each study, researchers document information on effect sizes and contextual factors (e.g., research design, subject populations) in spreadsheets, sometimes called evidence tables.
The data collected in evidence tables is then statistically aggregated in meta-analysis.
Following these procedures helps researchers answer a targeted research question while to mitigating potential biases (e.g., selection bias, confirmation bias) that might otherwise accrue.

In practice, researchers sometimes do not adhere strictly to these standards.
Researchers under time pressure may take shortcuts (i.e., rapid review~\cite{Ganann2010,Khangura2012,Watt2008}) by surveying the literature through citation trails or by making inclusion and exclusion decisions individually.  
If the literature on a topic is sparse or studies cover a variety of populations and contexts,
it is difficult to follow conventions of systematic review and conventional meta-analysis may be inadvisable~\cite{Cooper2009,Lipsey2001}. Researchers may choose to conduct a scoping study~\cite{Arksey2005,Levac2010} in which they survey the breadth of literature to identify gaps in knowledge.
The impacts of rapid review and scoping study methods on quality of findings have only been studied recently~\cite{Pham2014,Tricco2015}, and there is disagreement about best practices.
We contribute a characterization of the gap between best practice and actual practice.

\subsection{Softare for Research Synthesis}
Interactive systems for research synthesis offer relatively little support for reasoning about possible analysis paths.
Software tends to provide features for common steps in the analysis process such as forms for risk of bias assessment~\cite{eppi-reviewer,revman}, data extraction~\cite{eppi-reviewer,revman}, or creating forest and funnel plots of meta-analytic results~\cite{cma,eppi-reviewer,metafor,mix,revman}.
Additionally, most tools focus exclusively on a single stage in the analysis process~\cite{Bax2007} such as study screening~\cite{rayyan} or meta-analysis~\cite{cma,metafor,mix}.
Designing research software as a set of isolated optional procedures forces the researcher to conduct alternative analyses sequentially and separately. 
Further, this design fails to represent the motivations and constraints that drive researchers' decision-making.

Three tools 
in particular---RevMan~\cite{revman}, Eppi-Reviewer~\cite{eppi-reviewer}, and Rayyan~\cite{rayyan}---
offer collaboration features, such as role assignment~\cite{revman}, review flow diagrams~\cite{eppi-reviewer}, and coding disagreement overviews~\cite{rayyan}, which help researchers coordinate and review their work.
However, with the exception of highlighting disagreements about study inclusion/exclusion (a decision with implicit alternatives), these features do not help researchers identify and weigh alternative analyses.
Eppi-Reviewer~\cite{eppi-reviewer} allows users to create custom ``codesets'' and annotations to code the status of studies under review and take notes.
However, researchers may overlook the benefits of using these features to document motivations and constraints that guide their analyses in order to support later recall or scrutiny, instead using them inconsistently or not at all.
Rayyan~\cite{rayyan} uses a pulldown interface to enable researchers to select or create a reason for their decision to include/exclude a study. 
While this elicitation technique is a promising way of linking reasoning to decisions, Rayyan does not document competing motivations or constraints unless they are provided by different users who disagree about study screening.

We point to opportunities for tools to better support researchers in exploring, reasoning about, and communicating uncertainty about alternative analyses.

\section{Method}
We conducted open-ended conversational interviews with professional researchers to investigate practices in applied research synthesis. 
The goals of these interviews were to (1) characterize how researchers manage possible analysis paths, (2) gather information about the reasoning behind their choices, and (3) study how interactive systems can support awareness of uncertainty in analytical decision-making. 

\begin{figure*}[ht]
    \begin{centering}
    \includegraphics[width=\textwidth]{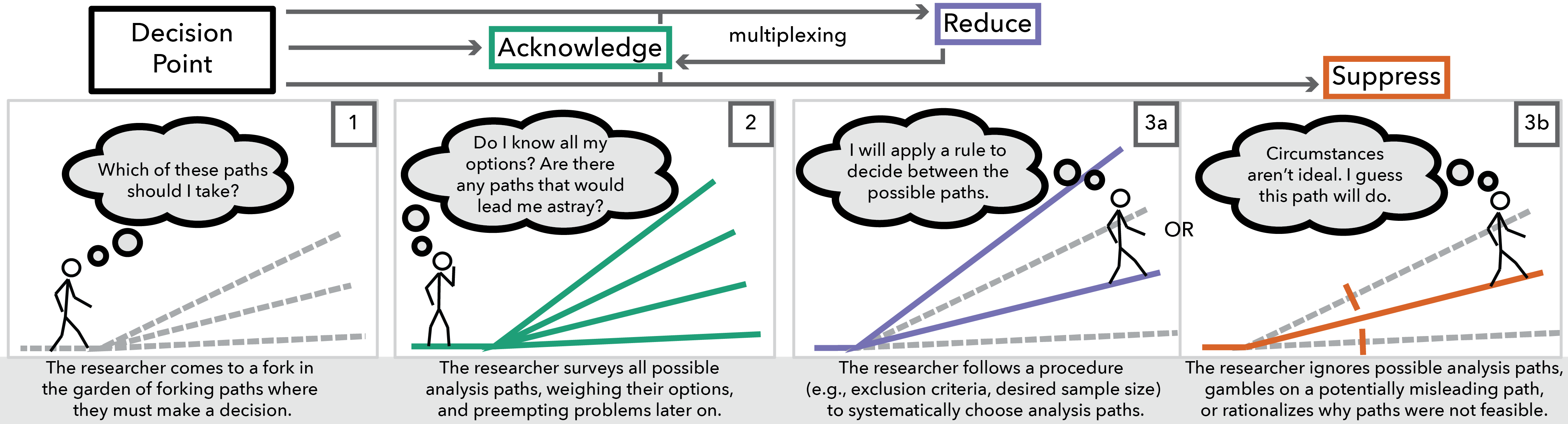}
    \end{centering}
    \caption{A pictorial representation of strategies used to navigate the garden of forking paths. The diagram at the top depicts observed temporal orderings of these strategies, e.g., multiplexing across choices in a loop of acknowledgement and reduction.
    }
    \Description{Read the following as a four pane comic with different possible orderings of the last three panes.
    Decision-point: ``Which of these paths should I take?'' The researcher comes to fork in the garden of forking paths where they must make a decision.
    Acknowledge: ``Do I know all my options? Are there any paths that would lead me astray?'' The researcher surveys all possible analysis paths, weighing their options, and preempting problems later on.
    Reduce: ``I will apply a rule to decide between the possible paths.'' The researcher follows a procedure (e.g., exclusion criteria, desired sample size) to systematically choose analysis paths.
    Suppress: ``Circumstances aren't ideal. I guess this path will do.'' The researcher ignores possible analysis paths, gambles on a potentially misleading path, or rationalizes why paths were not feasible.
    }
    \label{fig:strategies}
\end{figure*}

\subsection{Sampling Participants}
We employed convenience and snowball sampling~\cite{Creswell2018} to find 11 professional scientists to interview for our study.
First, we reached out to four professors at the Medical Center of a large public university who had recently published at least one systematic review or meta-analysis to advocate for evidence-based practices, such as utilitarian healthcare policies. 
Through these professors, we were able to interview one additional postdoctoral researcher studying teaching strategies in STEM. 
Second, we recruited two PhDs from the Veterans' Affairs Medical Center in a major US city working on systematic review and meta-analysis to recommend treatments for veterans with post-traumatic stress disorder (PTSD) and substance use disorder (SUD). 
Last, we used connections in the Navy to recruit four scientists using research synthesis to recommend improvements in training practices for military pilots. 
Of all the 13 people we contacted, only two declined to interview.
Our sample represents people doing research synthesis in formal professional settings.

\subsection{Interview Guide}
We created an open-ended conversational interview guide~\cite{Lofland2006,Miles2014} with the objective of getting participants to discuss their research practices in terms of specific examples.
The interview guide was a list of topics of interest regarding decision-points at different stages of the research process: scoping research questions, sampling literature, assessing and organizing evidence, analysis and visualization, and communicating findings (see Supplemental Material~\footnote{\url{https://github.com/kalealex/analysis_paths_research_synthesis/tree/master/interview}}). For the first five interviews, the guide was formatted as a list of questions, but we abridged these questions to a list of topics~\cite{Lofland2006} covering the same content in order to better accommodate the need to ask questions in terms of the experiences of individual researchers~\cite{Miles2014}, whose work varied in methods and settings.
Most often the interviewer broached a topic by asking a question of the form, ``How did you...''
For example, ``When you conducted that literature review, how did you decide which papers to include or exclude from your review?'' 
The interviewer also asked follow-up questions to seek clarification or greater detail about particular analytical decisions. 
The interview guide structured our conversations around a consistent set of topics while allowing flexibility to probe for a greater depth of description when necessary.

\subsection{Qualitative Coding Process}
All interviews, transcription, qualitative coding, and analysis were conducted by the first author, with iterative feedback on the coding and analysis framework from the other two authors. The analyses presented in this paper represent the perspectives of our participants systematically curated in an interpretive framework which was developed through discussions among the authors.

\subsubsection{Transcription}
The first author listened to audio recordings of interviews and transcribed all episodes of interest, omitting from transcription only small talk that was obviously not relevant to research synthesis. 

\subsubsection{Creating a Coding Framework}


To build familiarity with the content in the early interviews, the first author used open coding~\cite{Creswell2018} to describe what participants said about their research practices (e.g., literature review, meta-analysis, communication). 
Open coding helped us determine what we could reasonably infer from our interviews, similar to the use of grounded theory in related prior work~\cite{Boukhelifa2017}. 
Three provisional themes emerged. 
Participants described \textit{decisions} they made throughout the scientific process, the \textit{reasons and constraints} which guided those decisions, and specific ideas about \textit{how software features could support their work}. 

We used the themes identified in open coding to develop a more targeted framework (see Analytical Framework) characterizing analytical decisions and the reasoning behind them. 
For each decision, we made a set of categorical judgments about the nature of the decision and the rationale provided and recorded contextual information about the practices described. 
We continued to code instances where interviewees stated needs for software support in order to help us identify design opportunities in software for research synthesis.
We kept track of these codes in a spreadsheet with one row per decision (see Supplemental Material~\footnote{\url{https://github.com/kalealex/analysis_paths_research_synthesis/tree/master/analysis}}).

Fine-tuning this framework 
was an iterative process.
The first author coded analytical decisions and noted issues that came up during coding.
Then, the first author presented these issues to collaborators for feedback and discussion. 
The final coding scheme (described below) represents our consensus about how to best characterize the practices described by our participants as they relate to the challenges of navigating the garden of forking paths.

\subsection{Analytical Framework}
At the heart of the framework are \textbf{analytical decisions} described by participants. 
We were primarily concerned with how each decision acted on the space of possible analysis paths, for example, by surveying multiple paths or selecting one path through some procedure (Fig.~\ref{fig:strategies}). 
Following Lipshitz and Strauss~\cite{Lipshitz1997}, we coded each decision as an instantiation of one of the following three strategies:
\begin{enumerate}
    \item \textit{Acknowledge [Ack.]} uncertainty by accounting for different possibilities and planning to confront or avoid potential risks. This includes decisions to explore possible analysis paths, orient the research toward a broad set of issues, check for potential problems, and provide details or caveats in order to preempt misinterpretations of scientific evidence.
    \item \textit{Reduce [Red.]} uncertainty by gathering information, applying rules or conventions, making assumptions, or adopting a specific procedure to exert control over uncertainty. This includes decisions to seek information on a particular topic, to select and encode specific evidence, 
    and to follow a rule (e.g., inclusion criteria) at a decision-point.
    \item \textit{Suppress [Sup.]} uncertainty by eliminating possibilities through intuition, guessing, 
    or other procedures which accumulate unnecessary error. We only coded decisions as suppression when participants expressed that an alternative analysis path would contribute less error to their analysis.
\end{enumerate}

These strategies entail different levels of justification for analytical decisions and tend to have a \textit{temporal order} (Fig.~\ref{fig:strategies}, top). 
Imagine a researcher conducting a meta-analysis on the effect of mindfulness on depression.
The researcher first decides to search the ``gray literature'' (i.e., unpublished or unconventional sources), orienting the scope of their search before engaging in targeted information retrieval, an example of the acknowledgement strategy.
Decisions which acknowledge uncertainty often steer research broadly, surveying possible analysis paths and associated trade-offs prior to definitive choices about how to implement the analysis. 
Next, while reviewing the gray literature for their meta-analysis, the researcher decides to search unpublished dissertations for relevant data, an example of the reduction strategy.
Decisions which reduce uncertainty use specific procedures to navigate analysis paths, sometimes pruning away possible paths by omission. 
Reduction often 
builds on acknowledgement by implementing the broad goals identified in acknowledgement through a specific approach to analysis.
While still engaged in data collection for their meta-analysis on mindfulness, the researcher chooses not to examine the risk of bias in individual studies in their sample. 
The researcher knows this is not ideal but rationalizes this decision because they have limited time to conduct their analysis.
Decisions which suppress uncertainty are often necessary but unjustifiable compromises in response to situational factors which are sometimes beyond the researcher's control. 
Suppression substitutes for acknowledgment or reduction strategies. 

We also coded aspects of decision context.
We recorded the \textbf{stage in the research process} (e.g., \textit{question formation, literature review, meta-analysis}) for each decision. 
This allowed us to compare the relative frequencies of different strategies at different stages in the research process, roughly following the way that Wicherts et al.~\cite{Wicherts2016} break up researcher degrees of freedom into different phases of analysis. 

We coded the \textbf{reason given for each decision} (e.g., striving for \textit{reproducibility, limited availability of information, standard practices}). 
Sometimes these reasons were not explicitly stated, and we had to infer them from the broader context of our discussions with participants. 
Coding the reasons for each decision allowed us to examine what factors motivated participants to engage in each strategy. 


We extended the notion of the garden of forking paths to \textbf{communicative and organizational decisions}. 
We coded a distinction between decisions which have a \textit{direct impact} on the final written report of results and decisions which have an \textit{indirect impact} on results through interpretive and communicative aspects of the scientific process.
This distinction allowed us to highlight analytical decisions with impacts that may be difficult or impossible to quantify.

We also coded \textbf{metadata} such as the \textit{participant}, their \textit{field of study}, and the \textit{goal} of their project. 
Lastly, we noted when a decision was associated with a particular \textit{need for software features} or a particular \textit{threat to validity}. 
These passages informed our discussion of 
challenges and opportunities in designing for the garden of forking paths.

\section{Results}

\begin{figure}[ht]
\begin{center}
    \includegraphics[]{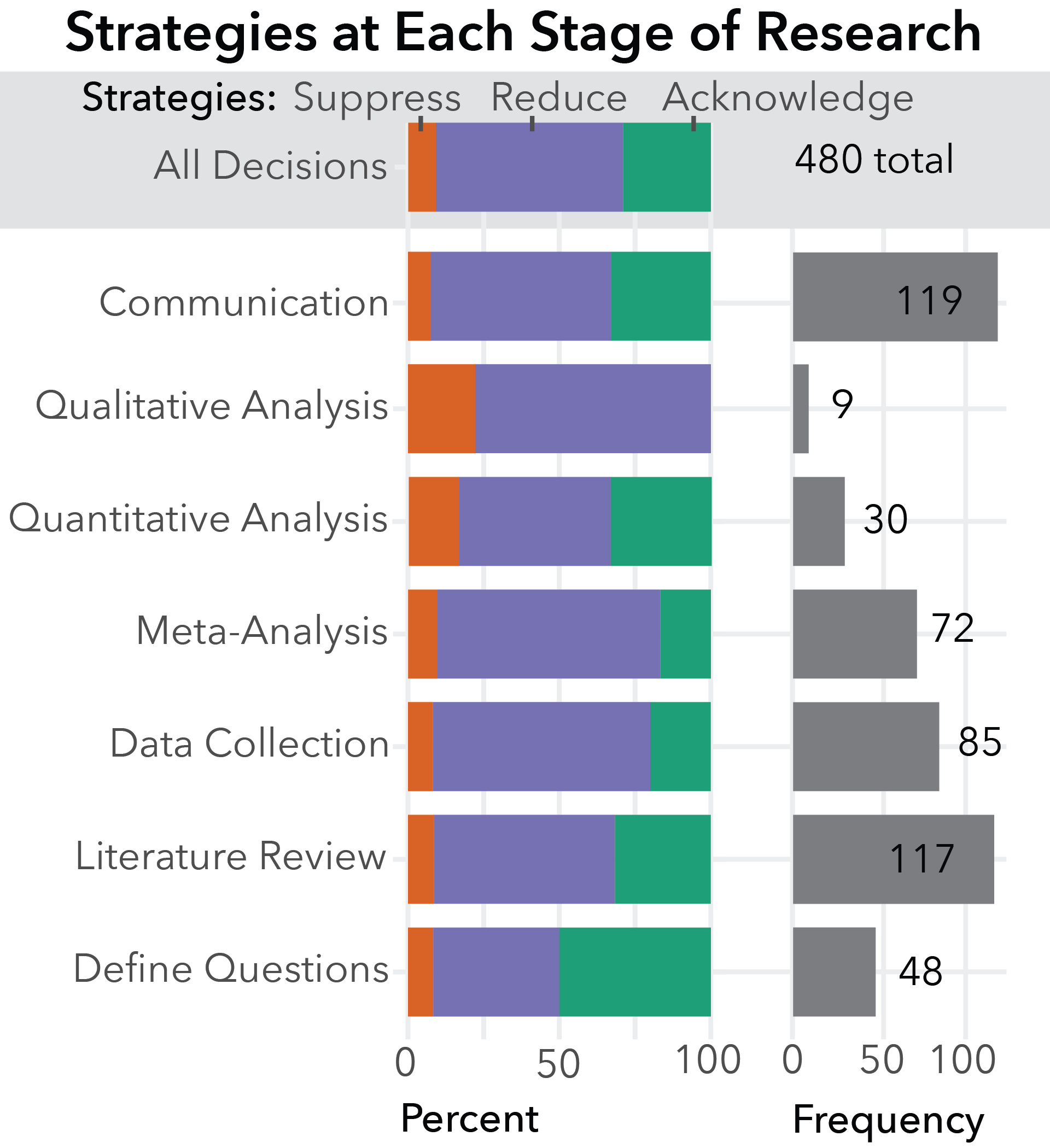}
\end{center}
 \caption{Frequencies of analytical decisions and percentages of strategies coded at each stage of the research process.}
 \Description{Statistics of interest are given in the text of the paper as well, but here's the dataframe containing the data depicted by this figure:
Strategy Frequency    Percent            Stage
1  Acknowledge       139  28.958333          Overall
2       Reduce       297  61.875000          Overall
3     Suppress        44   9.166667          Overall
4  Acknowledge        24  50.000000 Define Questions
5       Reduce        20  41.666667 Define Questions
6     Suppress         4   8.333333 Define Questions
7          All        48 100.000000 Define Questions
8  Acknowledge        37  31.623932       Lit Review
9       Reduce        70  59.829060       Lit Review
10    Suppress        10   8.547009       Lit Review
11         All       117 100.000000       Lit Review
12 Acknowledge        17  20.000000  Data Collection
13      Reduce        61  71.764706  Data Collection
14    Suppress         7   8.235294  Data Collection
15         All        85 100.000000  Data Collection
16 Acknowledge        12  16.666667    Meta-Analysis
17      Reduce        53  73.611111    Meta-Analysis
18    Suppress         7   9.722222    Meta-Analysis
19         All        72 100.000000    Meta-Analysis
20 Acknowledge        10  33.333333   Quant Analysis
21      Reduce        15  50.000000   Quant Analysis
22    Suppress         5  16.666667   Quant Analysis
23         All        30 100.000000   Quant Analysis
24 Acknowledge         0   0.000000    Qual Analysis
25      Reduce         7  77.777778    Qual Analysis
26    Suppress         2  22.222222    Qual Analysis
27         All         9 100.000000    Qual Analysis
28 Acknowledge        39  32.773109    Communication
29      Reduce        71  59.663866    Communication
30    Suppress         9   7.563025    Communication
31         All       119 100.000000    Communication}
 \label{fig:stages}
 \vspace{-5mm}
\end{figure}

\subsection{Decision-Making Strategies}

We coded 480 analytical decisions in our interviews. The majority were decisions to reduce uncertainty (297 of 480 decisions; 61.9\%), followed by decisions to acknowledge uncertainty (139 of 480 decisions; 29.0\%) and suppress uncertainty (44 of 480 decisions; 9.2\%).

\subsubsection{Acknowledgement}
\textit{Acknowledging uncertainty} was most prevalent early in the research process when researchers define questions and objectives (24 of 48 decisions; 50.0\%) (Fig.~\ref{fig:stages}). 
Participants identified this early scoping of the review as very important to a straightforward and systematic analysis. 
\textit{``If I spend time thinking about and writing down what my population, intervention, etc. is for my question, then that makes my literature review that much more efficient. Because then, as I run across a new study, it should be that much easier to say, `Is it in or is it out? Does it give me one of my outcomes that I've specified?' If it doesn't, then it's not included. It prevents you from getting mired down in indecision.'' [Ack.]} (P3).
Multiple researchers described a similar practice of considering possible analysis paths early on and writing a scope description to guide later analytical decisions.
Although researchers often rely on frameworks (e.g., PICOTS, described by Nelson~\cite{Nelson2014}) to guide scope development, links between scope and subsequent decisions are maintained in digital or paper notes if they are documented at all. 

We also see relatively more \textit{acknowledgement} during literature review (37 of 117 decisions; 31.6\%), quantitative analysis (10 of 30 decisions; 33.3\%), and communication (39 or 119 decisions; 32.8\%) than in data collection (17 of 85 decisions; 20.0\%) and meta-analysis (12 of 72 decisions; 16.7\%) (Fig.~\ref{fig:stages}). 
Researchers acknowledge uncertainty in later stages of research mostly to confront irreducible sources of uncertainty.
For example, one researcher in the Navy described using caveats to qualify information gathered through interviews with Navy personnel.
\textit{``Usually it ends up being a time issue or access. [We] did not have time to go out and validate that this is how they actually perform the work on the job. We just take people's word for it.'' [Ack.]} (P2).
In a similar case of acknowledgement to avoid potential misunderstandings of evidence, another researcher described checking the quality of available data.
\textit{``I'll use data visualization as I'm going through the process to see if there are things that look really weird. Like, I'll make a little ggplot and look to see if things are falling in the range that I expect them to.'' [Ack.]} (P9). 
When available information or resources constrain analysis paths, researchers tend to rely on acknowledgement to communicate limitations or check impacts on data quality.


\subsubsection{Reduction vs Suppression}
The high frequency of strategies to \textit{reduce} uncertainty at every stage in analysis (Fig.~\ref{fig:stages}) suggests that researchers often employ rule-based reasoning when implementing their scientific review and analysis. 
These rules are a mix of standard practices and lab-specific procedures.
For example, one researcher studying behavioral treatments for PTSD augmented the common practice of meta-analyzing only between-subjects experiments by looking separately at within-subjects evaluations to validate the treatment effect within an individual.
\textit{``Most meta-analyses only look at the between-group [studies], and they just assume that there is sufficient within-subject change to warrant doing any of it. I don't think that is as useful. That's the other big thing we are adding with this meta-analysis are these within-subjects tests to contextualize the between-group [effect]. I've never seen a meta-analysis in my area that does both, but systematic reviews do.'' [Red.]} (P10).
All researchers expressed awareness of best practices and deviated from them at times, but researchers had different attitudes about when and how much it was appropriate to bend the rules.

Sometimes researchers have little choice but to follow a path which they know is not ideal. 
For example, a Navy scientist measuring in-flight blood-oxygen levels to fill a gap in existing evidence described a decision to use finger-mounted monitors rather than more precise head-mounted monitors.
\textit{``Constraints of the experiment make it so that you can't get big head-mounted monitors and have a pilot wear their helmet at the same time. You have to make concessions where you can and realize that there's going to be some variability and error in your data just based on where your monitors are at.'' [Sup.]} (P5). 
The participant went on to describe how they would acknowledge this measurement error in their written report. 
When the best option available to researchers adds a source of error to analysis, the line between strategies of reduction and suppression is blurred, and researchers need to document how their analysis is constrained in order to alert stakeholders to potential \textit{suppression} of uncertainty.

Although we code \textit{suppression of uncertainty} infrequently (44 of 480 decisions; 9.2\%), we suspect that suppression is underrepresented in our sample 
because researchers may not recognize or admit when decisions introduce greater error than other viable alternatives.
For example, \textit{``We are for sure not including any qualitative research. The only two outcome measures we are interested in are exam score and failure rate. So maybe active learning changed how students feel about [the classroom climate], and maybe they have qualitative data from a survey. We are not including that either, and that's driven completely by our research question.'' [Red.]} (P4).
Although the researcher attributed the decision to ignore qualitative work to their research question, earlier in our interview they described potential biases in the framing of their research question.
\textit{``Before we started this meta-analysis, we wondered if maybe class size would be an explanatory variable in how well active learning works, or subject area, or whether it's an intro class or an upper-division class. Things like that, our relatively small research team came up with those things, so there's totally a bias present in the things that we omitted and the things that we included.'' [Sup.]} (P4).
Researchers need ways to represent and keep track of the reasoning behind analytical decisions in part because, as this researcher put it, \textit{``Defining where the personal bias is coming in and where specifically it's problematic would be important.''} (P4).
We argue that helping researchers document sources of bias, and how they influence analytical decisions throughout a research project, would serve to identify situational sources of uncertainty which are opaque in current research practice.

\subsection{Communicative and Organizational Practices}


We made a distinction between decisions which have \textit{direct impact} on the content presented in the written report of findings and decisions which have \textit{indirect impact} on findings.
Decisions with indirect impact do not change the quantitative or qualitative evidence presented in the written report of findings, but they change the way 
evidence is recorded in work documents and framed in meetings with collaborators or presentations to stakeholders. 
These social and communicative aspects of the research process have the potential to impact how findings and recommendations are interpreted.

Decisions with \textit{indirect impact} rely more on \textit{acknowledging} uncertainty (79 of 201 decisions; 39.3\%) than decisions with direct impact (60 of 279 decisions; 21.5\%).
Often these are decisions about how to organize or manage a review.
\textit{``We create a timeline and meet once a week to share what we've learned... and see where we have some challenges that we need to work on. If we've got to jump in and help someone else, then we'll do that too.'' [Ack.]} (P6).
Decisions to discuss issues with collaborators often serve to acknowledge possible ways of handling subsequent decisions with direct impact, such as how to sample the literature.
\textit{``When the [search] term isn't clear, it's a nightmare... 
That's sort of this ongoing iterative process, and honestly I don't think there's a way around it... The conversations with collaborators have been the best way to [narrow the scope] in my experience.'' [Ack.]} (P9).

Other times, decisions with \textit{indirect impact} are about checking the evidence in the review to preempt potential problems later on.
\textit{``Before the literature search is a tenth over, you should revisit your template against the literature you've examined. You can double-check your template [for the evidence table] to make sure it's right, and you adjust it.'' [Ack.]} (P7). 
By checking that the columns of the evidence table adequately reflect important themes in the literature, the researcher ensures that their review is comprehensive.
Decisions with indirect impact are scattered across notes, spreadsheets, and personal correspondences, yet they play an important role in shaping the scientific process behind the scenes.

In contrast, decisions with \textit{direct impact} tend to rely more on the \textit{reduction} strategy (190 of 279 decisions; 68.1\%) than decisions with indirect impact (107 of 201 decisions; 53.2\%).
For example, one participant described a relaxed approach to systematic review which did not rely on evidence tables.
\textit{``We were kind of looking for whether a study was really sound or not. Some were more easily brushed aside because it seemed like their research design was not something we would have followed, so we trusted each other to weed some of those out early on.'' [Red.]} (P6).
This researcher's team excluded and reviewed studies outside the context of a consistent template.
When asked if they would use an evidence table, this participant said, \textit{``If there was a spreadsheet already created, and I could tailor it to what might be most important for our particular project, then I think I would use that. But as far as using it every time and pulling out every piece of [information], I would literally have to have an intern do that for me because I don't have time.''} (P6).
Shortcuts, such as forgoing dual-review and evidence tables, are essential to researchers in the Navy because they operate on tight timelines.

\begin{figure}[ht]
\begin{center}
    \includegraphics[]{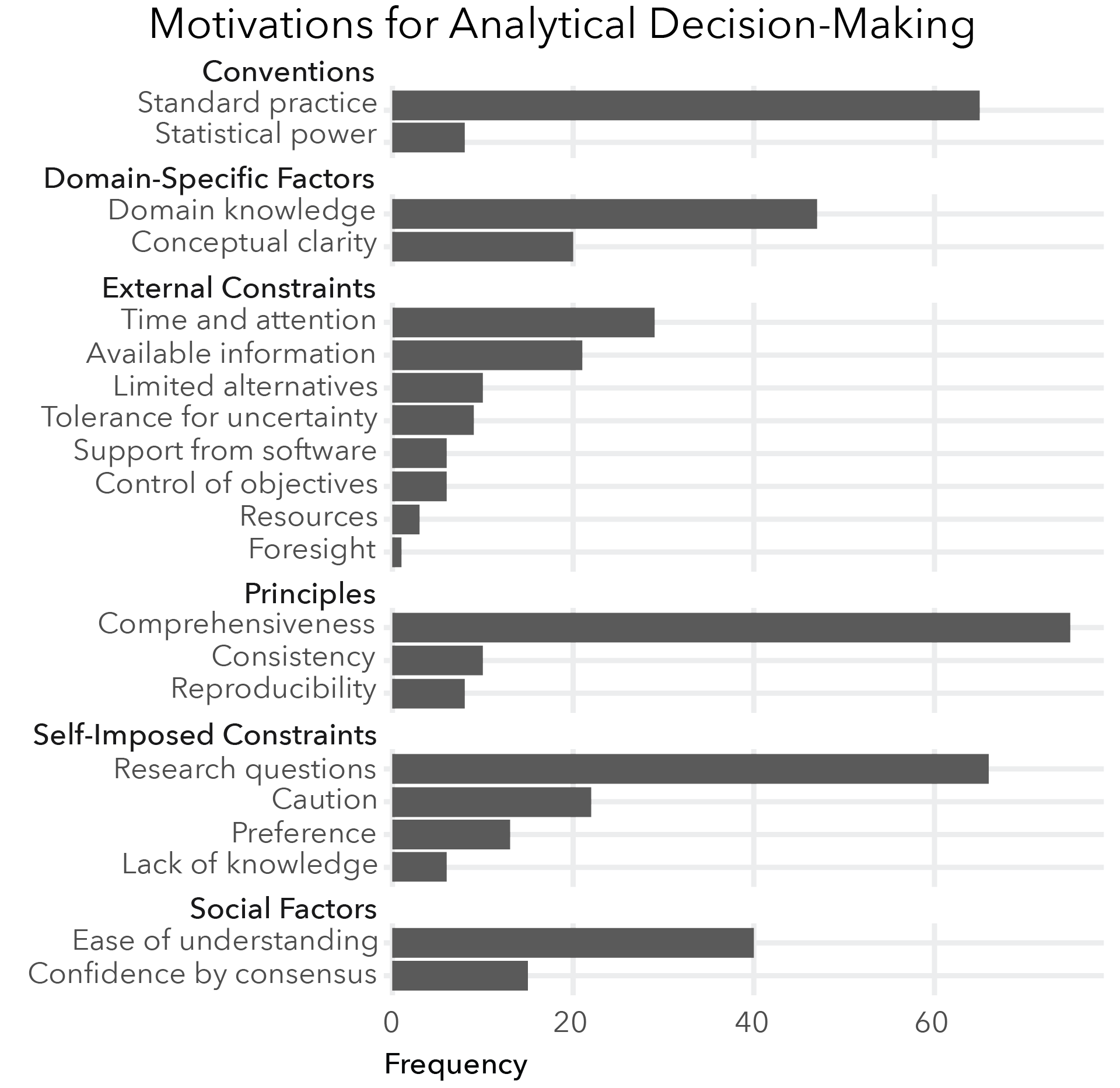}
\end{center}
 \caption{Categorized reasons for analytical decisions.}
 \Description{Statistics of interest are given in the text of the paper as well, but here are the dataframes containing the data depicted by this figure.
 Frequencies of reasons for decisions:
Reason count
1                                                  Caution    22
2                                        Comprehensiveness    75
3                                       Conceptual clarity    20
4                             Confidence through consensus    15
5                                              Consistency    10
6                                         Domain knowledge    47
7                                    Ease of understanding    40
8                                        Lack of knowledge     6
9                                     Limited alternatives    10
10                     Limited availability of information    21
11                  Limited control of research objectives     6
12                                       Limited foresight     1
13                                       Limited resources     3
14                     Limited support from software tools     6
15                              Limited time and attention    29
16 Limited tolerance for uncertainty or scientific process     9
17                                              Preference    13
18                                         Reproducibility     8
19                                      Research questions    66
20                                       Standard practice    65
21                                       Statistical power     8
 Frequences of motivations for decisions:
Motivation Frequency  Percent
1               Principles        93 19.37500
2           Social Factors        55 11.45833
3  Domain-Specific Factors        67 13.95833
4              Conventions        73 15.20833
5 Self-Imposed Constraints       107 22.29167
6     External Constraints        85 17.70833
 }
 \label{fig:motivations}
 \vspace{-5mm}
\end{figure}

\subsection{Motivations}

During our qualitative analysis we inductively coded 21 different reasons for researchers' decisions. We then grouped these 21 reasons into six themes which capture the motivations behind researchers' decision-making (Fig.~\ref{fig:motivations}). 

\begin{itemize}
    \item \textbf{Principles} (93 of 480 decisions; 19.4\%): ideals which researchers adhere to such as \textit{comprehensiveness} of the review,  \textit{reproducibility} in judgments, \textit{consistency} in how evidence is evaluated.
    \item \textbf{Social Factors} (55 of 480 decisions; 11.5\%): strictly communicative or interpersonal influences on decision-making such as a desire for \textit{ease of understanding} or building \textit{confidence through consensus}.
    \item \textbf{Domain-Specific Factors} (67 of 480 decisions; 14.0\%): concerns related to the domain under study such as cases where the researcher relies on \textit{domain knowledge} or seeks \textit{conceptual clarity} regarding key constructs.
    \item \textbf{Conventions} (73 of 480 decisions; 15.2\%): \textit{standard practices} which often reflect established knowledge about how to deal with issues like \textit{statistical power}.
    \item \textbf{Self-Imposed Constraints} (107 of 480 decisions; 22.3\%): factors which are under the control of the individual researcher such as \textit{research questions}, \textit{preferences}, or acting on a sense of \textit{caution} or a \textit{lack of knowledge} about research procedures.
    \item \textbf{External Constraints} (85 of 480 decisions; 17.7\%): factors beyond the researchers' control such as \textit{limited availability of information}, \textit{limited time and attention} for the review, \textit{limited control of research objectives}, or \textit{limited alternatives} to the chosen analysis path.
\end{itemize}

In our interviews, decisions to \textit{acknowledge} uncertainty were most often motivated by principles (46 of 139 decisions; 33.1\%), domain-specific factors (32 of 139 decisions; 23.0\%), and self-imposed constraints (21 of 139 decisions; 15.1\%). 
This suggests that acknowledging uncertainty is often voluntary and carefully thought out on the part of researchers. 
Researchers are particularly motivated by the principle of comprehensiveness (75 of 93 decisions; 80.6\%), often reporting that they provide detailed background information in their written report, check data quality, and code possible covariates in their evidence tables. These practices serve the purpose of a complete and thorough review.

The motivations for decisions to \textit{reduce} uncertainty are fairly balanced, with self-imposed constraints (81 of 297 decisions; 27.3\%) and conventions (60 of 297 decisions; 20.2\%) being most prevalent. 
By analyzing participants' motivations, we find that not all decisions to reduce uncertainty are equally justified, and this is further evidence of a blurred line between reduction and suppression strategies.
For example, one participant described flexibility in defining the search terms used to sample the literature.
\textit{``It's helpful when you have a sense of how many studies you're talking about. In the psychiatric conditions, there was a study looking at anger. Anger is not exactly---there's not a diagnosis associated with that. Initially, I wanted to include it, but there's only a couple of studies for which that is relevant, so it seemed like it was cleaner to just exclude anger because it's not a diagnosis. It's just two studies; it's not a big deal... We don't want to do a systematic review with two studies; we don't want to do a systematic review with 100 studies because it's probably not targeted enough. It seemed like some of that process was finding a happy medium.'' [Red.]} (P9). 
By sampling in order to hit a target sample size, the researcher may ignore constructs which are theoretically important but hard to measure.

In contrast, another researcher described a decision to \textit{include a group with two studies in a hierarchical meta-analysis} (P3) (i.e., meta-regression) in order to address their research question.
Later, when speaking in general terms, they said that using groups of studies which are too small in meta-regression leads to unrepresentative results and poor statistical power.
These decisions about how to group studies highlight a trade-off between conventions---adhering to statistical best practices by having large enough samples in each group---and self-imposed constraints---evaluating a specific research question by using a certain grouping factor in a statistical model. 
The frequency of external constraints as a motivator for suppression (33 of 44 decisions; 75.0\%) and the thin line between suppression and reduction strategies suggest that researchers would benefit from documenting the reasoning behind their analytical decisions.

\section{Design for the Garden of Forking Paths}
Our findings point to challenges in designing interactive systems to support research synthesis which faithfully represent the relevant empirical evidence while accounting for organizational and individual constraints and goals in the synthesis process. Specific challenges that emerge from our results include: 1) a tension in current practice between researchers' strong desire to consider multiple possible analysis paths, as evidenced by the prevalence of acknowledgement, and the frequent influence of self-imposed constraints in motivating the reduction of uncertainty;
2) a lack of consistent representations and processes for capturing their rationales for decisions, limiting researchers' ability to reflect on their rationales; and 3) a trade-off between researchers' desire to acknowledge uncertainty (e.g., by stating caveats) and the need to tell a clear story with their analysis (e.g., filtering evidence based on their research question).


\subsection{Research Synthesis as Expected Utility Maximization}
The inherent subjectivity in scoping a systematic review or meta-analysis, combined with the influence of external constraints, suggests that the single ``best'' analysis may not always be the analysis that is chosen in an organization. To guide reflection on how interactive software might better support researchers' endeavors, we frame research synthesis as a process aimed at \textit{maximizing the expected utility of analysis} as perceived by the researcher within an organization.

An expected utility framework assumes that a person will decide between multiple actions by reasoning about the expected consequences
of each in combination with the 
value they place on those consequences,
which is defined by a utility function~\cite{vonNeumann1944}.
While the consequences of actions (e.g., betting strategies) in a classic example of utility are monetary payoffs, we suggest that an adaption of 
utility theory for research synthesis might frame the relevant consequence of an action, or a choice of analysis,
as the accuracy or faithfulness with which the analysis captures the true state of the world.
Since the true state of the world is unknown, there is uncertainty about the faithfulness of any analysis. In research synthesis perceived uncertainty takes the form of the researcher's subjective sense of confidence in the accuracy of possible analyses.
The criterion for choosing between actions is the average utility of each action given the uncertainty.
Under utility theory a decision-maker should be able to choose the higher utility alternative given any pair of actions (or alternatively, should be able to declare them equal in utility)~\cite{jensen1967,vonNeumann1944}. 
This assumption implies that interactive systems for research synthesis could help researchers choose the paths that maximize their expected utility by helping them to 1) accurately perceive the consequences of different analysis paths (i.e., actions). 
Because the true state of the world, and thus the exact consequences of an analysis, may be impossible to judge, researchers may use heuristics such as looking at the results of different paths or considering competing values to make decisions.
Similarly, interactive systems could also support researchers directly in 2) identifying how much they value the accuracy of different paths.

\subsection{Acknowledging Alternative Analyses}

\subsubsection{Challenge: Trying a set of analyses instead of just one}
Our interviews suggest that one of the core challenges in designing for research synthesis is supporting judgments about when a given analysis is justified. The pitfall for researchers is the thin line between exploratory analysis and using researcher degrees of freedom to select an analysis path which confirms a particular hypothesis or point of view.
An episode from one of our interviews illustrates this ambiguity:
\textit{``Through those discussions that we have, we'll have suggestions for different types of analysis we can conduct that might split the hair a little bit differently and give us additional information that we can then decide, `Okay, what is the best way to report this now that we've looked at it both of these ways?'''} (P6). 
In cases like this, we argue that
any reduction strategy used to select a particular analysis may actually be suppressing uncertainty because presenting one analysis path among multiple paths observed may reflect confirmation bias as much as the true state of the world. 

Prior work on reliable statistics~\cite{Simonsohn2015,Steegen2016} suggests that researchers should survey possible analysis paths and multiplex across them by exploring \textit{and reporting} a reasonable subset of analyses.
Because comprehensiveness was the most prevalent motivation for analytical decisions in our analysis, we suspect that researchers are naturally incentivized to explore possible analyses and that they will do so to the extent that tools make it feasible.
Being able to directly evaluate how different paths impact results would provide researchers with relevant information for determining what paths have higher utility given their various motivations and constraints. For example, when the differences in results between paths are minimal, the researcher may decide to choose a path that ``trades-off'' accuracy with other desirable properties like minimizing time or resources spent, or maximizing consensus or interpretability~\footnote{An alternative framing of decisions about paths in our study results could define utility as a function of path accuracy alone, and choices of non-optimal paths as ``satisficing'' under bounded rationality~\cite{simon1972}}.
When results are highly variable, the researcher should exercise extreme caution and consider reporting multiple analyses if possible.

Of course, researchers may not always be motivated report on multiple analyses paths if they prefer a certain result, raising the ethical dilemma that developing tools which make it easier to explore alternative analyses may facilitate the cherry-picking of results.
The need to explore multiple analyses and the risk that this multiplexing will facilitate cherry-picking remains a tension in the broader set of recommendations in the literature on reducing bias in analysis.

\subsubsection{Opportunity: Multiverse analysis}
Each possible analysis in the garden of forking paths produces a distribution of estimates which represents error in the analysis process. This is distinct from uncertainty about which path most accurately represents the evidence (i.e., the true underlying effect).
Prior work~\cite{Simonsohn2015,Steegen2016} suggests that, ideally, researchers should quantify and report uncertainty about analytical decisions by running a subset of possible analyses to see which decisions impact results, a procedure called multiverse analysis. 
Similar work on model comparison techniques (e.g.,~\cite{Manski2003,Manski2018-lure,Piironen2017}) suggests that researchers should build and compare multiple models expressing assumptions of varying strength in order to separate uncertainty of evidence from uncertainty of assumptions in their analysis.
These convergent lines of research suggest an opportunity for interactive systems to elicit a set of analysis paths under consideration 
and support researchers in interactively comparing outcomes from multiple quantitative analyses.

\subsubsection{Opportunity: Visualizing the garden of forking paths}
In order to support researchers in exploring, comparing, and reporting on multiple possible analyses, interactive systems need an explicit way to represent analytical decisions and elicit information about possible analyses from researchers. 
Based on prior work on interactive visualization of scientific workflows~\cite{callahan2006vistrails}, we point out the opportunity to represent the garden of forking paths using interactive diagrams which help researchers map out decision-points and their influence on one another as nodes and edges.
We propose other opportunities related to reasoning with and communicating uncertainty which build on this representation. 

\subsection{Representing Reasoning About Analysis}

\subsubsection{Challenge: Shifting attention from rationales to impacts}
When prompted to describe the reasoning behind analytical decisions, researchers tend to focus on factors which rationalize their choices, often appealing to standard practices and research questions rather than examining the impacts of their choices on the results of analysis. 
For instance, we compare two cases where researchers made opposite decisions about whether or not to include a group of two studies in a meta-regression (see Results: Motivations).  
Both researchers were aware that two studies would not give them enough statistical power to make a precise effect size estimate, 
but instead they rationalized their decisions by appealing to conventions (P9) and research questions (P3).
We observe that researchers document their rationales inconsistently across personal correspondences, notes, and work documents. This may contribute to difficulty weighing these motivations alongside the outcomes of alternative analysis paths.
Considering rationales in the absence of information about outcomes leads to a sense of utility that is driven primarily by the perceived value of a choice independent of its consequences.
We argue that this problem would be mitigated by interactive systems that formally represent researchers' reasoning about analysis paths and attempt to shift researchers' attention to the impacts of decisions on the results of analysis. 

\subsubsection{Opportunity: Attributing rationale}
While researchers often have a rationale for their decisions, they seem to lack the tools to externalize and review the motivations and constraints which shape their analysis.
For example, one researcher expressed the need for a way to track conceptual uncertainty when coding evidence tables in Excel.
\textit{``If [a flag or annotation is] associated with the value in a cell, I think that would help coders feel more confident even if you're not ultimately going to do anything about it. And it might help your quality control checks at the end.''} (P4).
In agreement with this researcher, prior work on how uncertainty is represented in visual analytics systems~\cite{Sacha2016} suggests that helping users maintain awareness of sources of uncertainty improves the calibration of confidence in the accuracy of the analysis. 
Given an interactive visualization of the garden of forking paths,
a system could enable researchers to create custom flags associated with factors which influence analytical decision-making such as rules, assumptions, and conventions, as well as guidelines (e.g., scope descriptions and research questions) and constraints (e.g., limited time and attention).
Researchers could use these flags to represent their reasoning at a given decision-point by attributing their choice to a set of rationales.
We speculate that mapping out rationales for alternative analyses might highlight trade-offs between different motivations and thus prompt researchers to reflect on and potentially update their decisions.

\subsubsection{Opportunity: Aligning subjective and statistical uncertainty.}
Since the true state of the world is unknown, researchers must rely on their subjective sense of the accuracy of analyses (i.e., confidence) to decide between alternative paths.
However, some of the rationales given by researchers (e.g., conventions) are essentially appeals to authority, suggesting that researchers sometimes feel uncomfortable with their ability to judge the consequences of alternative analyses.
Prior work~\cite{hullman2018-imagining-replications} shows that comparing predicted effect size distributions to observed outcomes helps some form expectations about effects that better align with statistical uncertainty. 
Expressing subjective expectations promotes more active reasoning~\cite{Kim2017,Kim2018-others-expectations},
and using visual representations 
offloads information from working memory~\cite{Cox1999,Natter2005}, freeing up attention for metacognitive reflection~\cite{Schraw2006} about how effects are understood.
This suggests
an opportunity to help researchers calibrate their expectations about the impacts
of analytical decisions through comparisons of 
subjective uncertainty, elicited by a system using graphical or other formats (e.g.,~\cite{goldstein2014,hullman2018-imagining-replications,Kim2017,Kim2018-others-expectations,kim2019}), and statistical uncertainty, achieved through computation.
Based on prior work~\cite{Chance2000}, we speculate that the experiential learning that occurs through such prediction can ultimately help foster confidence as well as accuracy in one's predictions.

\subsection{Communicating Uncertainty}

\subsubsection{Challenge: Varying tolerance for uncertainty}
Researchers we interviewed tend to remain skeptical about their findings, but they often need to present findings in a way that offers convincing support for recommendations.
Contributing to this tension, it has been argued that decision-makers~\cite{Bar-hillel1993,Bonaccio2006,Budescu2000,Manski2018,Swol2005,Yaniv1995,Yaniv1997} (and people in general~\cite{Curley1985,Einhorn1985,Gardenfors1983}) have limited tolerance for uncertainty.  
\textit{``They want a visualization that shows gross effect, it gets down to the point, so they are no longer wanting a string of visualizations to illustrate every point. If you write about the methodology, you write about your data crunching, and then you show your visualization for your final result, that is what decision-makers are hungry for and are expecting now... They do care about the fidelity of the data, they just don't want a chart on it.''} (P11).
In practice, this Navy researcher told a simplified and compelling story advocating for a promising training program.
They emphasized the trade-off between downplaying uncertainty to get decisions made and doing just the opposite when there was concern about the safety of Navy personnel.
\textit{``If it's a really important decision that you need to impress upon them, that there is a significant thing to consider, then I will sometimes do the findings a disservice by presenting the problem and burying the potential compelling use case or storyline.''} (P11).

\subsubsection{Opportunity: Visualizing possible outcomes}
We argue that interactive systems for research synthesis should provide a set of visualizations for distributions of possible outcomes, which attempt to alleviate specific aversions that decision-makers have toward uncertainty.
One major aversion to uncertainty is that many people, even experts~\cite{belia2005,soyer2012}, find it hard to understand~\cite{Kahneman2011}.
Prior work in psychology~\cite{hasher1984,hertwig2004}, statistical reasoning~\cite{gigerenzer1995,goldstein2014,hoffrage1998}, and data visualization~\cite{fernandes2018,hullman2015,hullman2018-imagining-replications,kale2019-hops-trends,kay2016bus} suggests a remedy to this problem:
people reason about uncertainty most accurately when it is framed as frequencies of events, rather than probabilities or summary statistics. 
As such, visualizations of \textit{quantitative uncertainty} should convey the possibility of multiple outcomes using 
frequency framing to circumvent misunderstandings of uncertainty.
When it is important to convey uncertainty in possible outcomes with high fidelity, hypothetical outcome plots~\cite{hullman2015,kale2019-hops-trends,kim2019} and quantile dotplots~\cite{fernandes2018,hullman2018-imagining-replications,kay2016bus} are valuable visualization formats.
By presenting discrete outcomes, these formats align with results from prior work in statistics pedagogy suggesting that people develop better statistical intuitions via simulations~\cite{chance1999,chance2004,cumming1998,jamie2002,soyer2012}.

In contrast, when decision-makers 
have low tolerance for uncertainty because they want a clear yes-or-no answer, visualizations could be designed to emphasize modal results while still displaying uncertainty, as supported by most static representations of distributions, including intervals. 
Additionally, propagating uncertainty in the effect distribution to derived measures that may be more closely aligned with the decision-maker's utility function, such as money or time saved, might help decision-makers appreciate uncertainty information for the purpose of making more informed decisions. 
\textit{``There are these situations where you face the moral quandary of having to tell a company that you wasted their time [on inconclusive results] or giving them information that isn't very useful to them [because the analysis misrepresents available evidence to give an exaggerated sense of certainty].''} (P5).
We argue that providing researchers with communication techniques which counteract decision-makers' specific aversions to uncertainty would mitigate the misconception that an uncertain result is not presentable and does not contain useful information.
Of course, the best way to identify the impact of the techniques we propose is to evaluate decision-making given different communication strategies.

\subsubsection{Opportunity: Conveying qualitative uncertainties}
In our interviews we find that researchers express \textit{qualitative} forms of uncertainty such as the assumptions and constraints behind their analysis by writing caveats in limitations sections or preparing supplemental presentation slides.
Descriptive accounts of uncertainty help
researchers align the expectations of decision-makers with the quality of available scientific evidence.
In the words of one researcher,
\textit{``There is no such thing as a perfect study, but maybe if there was some way to make that information more readily available to researchers so that I clearly see a trend in limitations that are going on in this area of research. How am I not going to let that affect mine, or how can I at least convey that skepticism to my potential employer so that they know what might impact their results?''} (P5).
Prior work~\cite{Manski2018,Manski2018-lure} suggests that uncertainty about assumptions is often overlooked by decision-makers, leading to an exaggerated sense of certainty about analytical results.
Drawing on the 
opportunities we identify for representing reasoning about analytical decisions, a system could 
use
rationales and their associations with decisions to create summary representations of assumptions, constraints, and limitations which should inform the decision-maker's sense of utility.
Future work should explore the design space for representing and communicating these qualitative uncertainties.

\subsection{Limitations}
Limitations of our procedure for sampling analytical decisions introduce imprecision into the frequencies of decision-making strategies presented in our results.
The sample of researchers we interviewed may not be representative of the broader population of researchers working in applied settings. 
Additionally, we rely on each participant's account of whether decisions are justified to inform our coding of suppression. 
While this means that our data reflect the perspectives of our participants rather than our opinions, it also means that suppression may be relatively underrepresented in our sample because researchers may feel disinclined to admit to choosing less-than-ideal analysis paths.

We draw on expected utility theory to provide a basis for identifying ``better'' ways of supporting research synthesis despite the inevitability of organizational constraints the process. However, our treatment is not intended to be a formal model. How to characterize research synthesis 
as it occurs in organizations using theories like expected utility maximization or bounded rationality is left to future work.

\section{Conclusion}
We present a qualitative analysis of how researchers conducting applied research synthesis navigate the garden of forking paths: a series of analytical decision-points, each of which has the potential to influence findings.
Based on our interviews and analysis, we identify a set of design challenges around making it more feasible for researchers to try and report multiple analyses, shifting researchers' attention from rationales for decisions to impacts of decisions, and supporting uncertainty communication techniques which address specific aversions to uncertainty among decision-makers.
Considering evidence from our interviews in light of prior work, we point out opportunities for interactive systems to support research synthesis by helping researchers map the garden of forking paths, document their reasoning about analysis paths, and effectively communicate uncertainties impacting their analytical decision-making.

\section{Acknowledgements}
This work was funded by US Navy STTR Contract N68335-17-C-0410 in partnership with Stottler Henke Assoc, and NSF award \#1749266.

\bibliographystyle{ACM-Reference-Format}


\end{document}